\documentclass{article}

\usepackage{arxiv}

\usepackage[utf8]{inputenc} 
\usepackage[T1]{fontenc}    
\usepackage{hyperref}       
\usepackage{url}            
\usepackage{booktabs}       
\usepackage{amsfonts}       
\usepackage{nicefrac}       
\usepackage{microtype}      
\usepackage{lipsum}
\usepackage{multirow}
\usepackage{amsmath}
\usepackage{inputenc}
\usepackage{tipa}
\usepackage{graphicx}
\let\ipa\textipa

\title{EfficientNet-Absolute Zero for Continuous Speech Keyword Spotting}

\author{
  Amir Mohammad Rostami\\
  Department of Computer Engineering \\
  Amirkabir University of Technology\\
  \texttt{a.m.rostami@aut.ac.ir} \\
   \And
 Ali karimi \\
  Department of Computer Engineering \\
  Tehran University\\
 \texttt{aliiikarimi@ut.ac.ir} \\
 \And
 Mohammad Ali Akhaee \\
  Department of Electrical Engineering\\
  Tehran University\\
    \texttt{akhaee@ut.ac.ir} \\
}

\begin{document}
\maketitle

\begin{abstract}
Keyword spotting is a process of finding some specific words or phrases in recorded speeches by computers. Deep neural network algorithms, as a powerful engine, can handle this problem if they are trained over an appropriate dataset. To this end, the football keyword dataset (FKD), as a new keyword spotting dataset in Persian, is collected with crowdsourcing. This dataset contains nearly 31000 samples in 18 classes. The continuous speech synthesis method proposed to made FKD usable in the practical application which works with continuous speeches. Besides, we proposed a lightweight architecture called EfficientNet-A0 (absolute zero) by applying the compound scaling method on EfficientNet-B0 for keyword spotting task. Finally, the proposed architecture is evaluated with various models. It is realized that EfficientNet-A0 and Resnet models outperform other models on this dataset.

\end{abstract}


\section{Introduction}

Keyword spotting in a continuous speech can be considered as a limited or particular case of large vocabulary continuous speech recognition (LVCSR) systems. Lightweight models that reduce the training and inference runtime, along with the growing usage of smartphones, has made this field applicable in business projects. Hey Siri, Hey Bixbi, and Hey Google, which is known as wake up or trigger systems are a particular mode of keyword spotting. In addition, home voice assistant systems are one of the keyword spotting applications.

Over the past several years, deep neural network-based methods have been used in various researches due to their remarkable performance and achievements. In order to use these methods in the keyword spotting field, it is necessary to have a sufficient and appropriate dataset along with an architecture that can be appropriately performed on smartphones or embedded systems. The dataset used for this purpose should have sufficient samples of each class or word from different aspects like word expression, accent, unique speakers, recording environment, recording systems, and ambient noise to improve the models' final performance in practical applications. On the other hand, to obtain excellent performance in continuous speeches, the training data must also contain this type of samples. The football keywords dataset (FKD) attempts to get all of these requirements in Persian language. This dataset is open-source\footnote{https://drive.google.com/file/d/1m0CoqVzneGVxfTx-uGpAXdZvPNgH9gjS/view?usp=sharing}for any research project. Determining the events of a football game using the reporter's speech is another reason for collecting this dataset. Our primary focus is on introducing the prepared dataset and examining the EfficientNet model \cite{eff} along with other methods on this dataset.

The rest of the paper is organized as follows. Section 2 is dealing with the related works and datasets in keyword spotting. Section 3 describes the details of preparing FKD and its features. In the next section, how the EfficientNet is used in this task will be introduced. In Section 5, the experiment configuration as well as the results have been described. Finally, Section 6 concludes the paper.

\begin{table}[h]\centering
\caption{Football keywords dataset classes with the international phonetic alphabet (IPA)}
\label{tbl:fkd_info}
\begin{tabular}{llccc}
\hline
\multirow{2}{*}{\textbf{Class}} & \multirow{2}{*}{\textbf{IPA}}  & \multicolumn{3}{c}{\textbf{Number of Utterances}}\\
  \cline{3-5}
                       &                       & Train         & Test        & CSS        \\
                       \hline
Corner                 & /korner/              & 1322          & 300         & 22         \\
Foul                   & /xatâ/                & 1369          & 300         & 3          \\
Free kick              & /kâ\ipa{S}teh/              & 1506          & 300         & 80         \\
Goal                   & /gol/                 & 1370          & 300         & 53         \\
Goalposts              & /tirak/               & 1376          & 300         & 91         \\
Hand                   & /hand/                & 1588          & 300         & 53         \\
Laying off             & /exrâj/               & 1354          & 300         & 72         \\
Mulct                  & /jarimeh/             & 1533          & 300         & 12         \\
Notice                 & /extâr/               & 1311          & 300         & 39         \\
Offside                & /âfsajd/              & 1268          & 300         & 56         \\
Out                    & /ot/                  & 1323          & 300         & 56         \\
Penalty                & /penâlti/             & 1375          & 300         & 91         \\
Red card               & /kârte qermez/        & 1219          & 300         & 30         \\
Strike                 & /zarbeh/              & 1358          & 300         & 93         \\
Substitute             & /ta?viz/              & 1298          & 300         & 81         \\
Tackle                 & /takl/                & 1245          & 300         & 34         \\
Throw-in               & /partâb/              & 1331          & 300         & 73         \\
Yellow card            & /kârte zard/          & 1289          & 300         & 23         \\
Silence                & \multicolumn{1}{c}{-} & -             & 300         & 20         \\
Unknown                & \multicolumn{1}{c}{-} & -             & 300         & 20         \\
\hline
\multicolumn{2}{l}{\textbf{Total}}  & \textbf{24435}         & \textbf{6000}        & \textbf{1002}\\
\hline
\end{tabular}
\end{table}
\section{Related works}
EfficientNet architecture is a powerful method that has achieved the best performance in recent years in many image processing tasks, especially on the Imagenet dataset. Moreover, the performance and efficiency of this architecture in speech recognition \cite{effspeech}, which is a more complicated task than keyword spotting, is significant and has attracted much attentions.

In this field, many works have employed shallow or deep models based on their purposes, resources, and trends of that time. Various researches have used convolutional neural networks (CNN) \cite{cnn2}\cite{cnn1}\cite{bbn}, but the main point is that this model has high computations and some works tried to solve this challenge with small-footprint architectures\cite{honk}\cite{cnnsmall}. For instance, Residual architecture methods that have the potential to go deeper have been employed in most signal processing areas such as image recognition, speaker identification/verification, speech recognition, and keyword spotting \cite{honk}. Feed-forward deep neural network (DNN) methods that paid attention to being small-footprint also used in this field\cite{DNN1}\cite{DNN2}\cite{DNN3}. However, these methods have lower accuracy and performance compared to convolution methods. In addition to these architectures, recurrent neural networks (RNN) are recently widespread in this field \cite{rnn2}. Although RNNs introduce some advantages, we do not consider them because of their high computational complexities and run times. Given these points, some works have also tried to combine convolutional and recurrent methods to achieve all benefits \cite{tfc}\cite{crnn}\cite{crnn2}. These methods performed well, but computational complexity remains their main challenge.

Besides deep neural networks, some researches are dedicated to applying several methods like hidden Markov models (HMM) \cite{hmm}, hybrid HMM-NN models \cite{hyb}, segmental Gaussian mixture models \cite{msr}, and minimum edit distance \cite{med}. However, these methods are less investigated these days due to deep neural networks' excellent performance which is achieved thanks to sufficient data and graphical processor units (GPUs).

Nowadays, a dataset published by Google has led to much progress in this field \cite{scd}. Google speech command (GSC) is available as an open-source for researchers and commercial works. In GSC, they considered most of the required phonemes, words of commands, and numbers. The dataset published by Mozilla \cite{mozila} has been one of the main factors in developing speech recognition systems because it is multilingualism and open-source. Deep mine corpus \cite{deepmine} is another Persian dataset released recently for text-dependent, text-independent, and text-prompted speaker identification/verification and speech recognition systems.
\begin{figure*}[t]
  \centering
  \includegraphics[width=\linewidth]{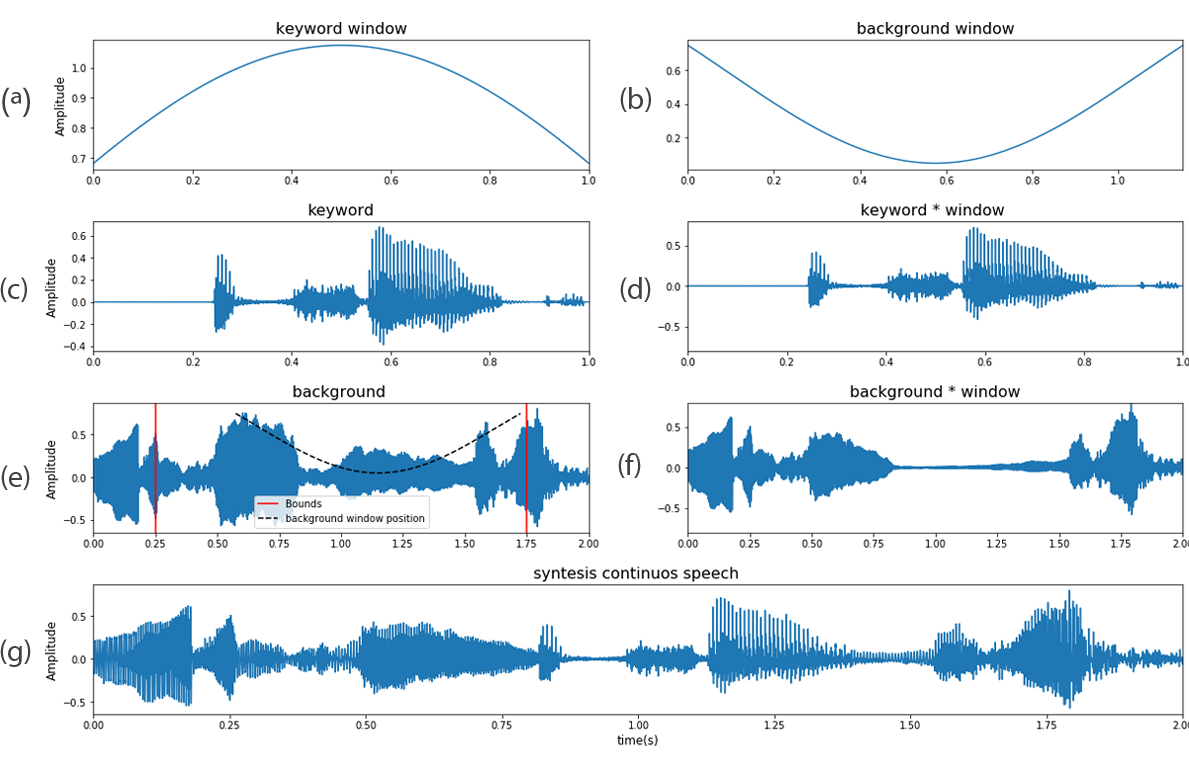}
  \caption{Steps of the continuous speech synthesis method. Windows that applied on the keyword (a) and background (b), sample signal waveform (c) and signal multiplied by the keyword window (d), sample background signal waveform (e) and position of the background window in left and background after multiply the window (f), and finally synthesis continuous speech sample (g).}
  \label{fig:cssm}
\end{figure*}
\section{Football Keywords Dataset }
\subsection{Main approach}
Detecting the events and actions of a football game and making an appropriate dataset for Persian keywords spotting in both single word and continuous mode (i.e., in the sentence) are our main reasons for preparing this dataset. In order to achieve this goal and obtain a generalized model, we consider the diversity of FKD from different aspects such as age, gender, mood, and tone of an utter, recording device, and recording environment in the collecting data phase.

The keywords used in this dataset are very close to the same words in English used in football. However, there exist some minor differences in pronunciation and accent. Words like goal, offside, corner, penalty, and hand in most languages have the same meaning and usage. Therefore, models employing English datasets can use this dataset for generalization and improvement of their systems.

The list of words is shown in Table~\ref{tbl:fkd_info}. We consider the majority of football events, which can be recognized through the football reporter's voice. On the other hand, during data collection, seven types of expression such as \textit{normal, emphasized, upset, surprised, emotional, fast,} and \textit{stretched} have been collected from the speakers so that the prepared dataset is comprehensiveness to some extent. This point, along with the variety of speakers, recording systems, and ambient noise, makes this dataset general and applicable in practical applications.

Since it is decided to collect the dataset from crowdsourcing, an attempt has been made not to ask the age, gender, and names of the participants. However, according to the collected speeches, the majority of utterances are men and young people. Fortunately, since the number of speakers is high, there exist enough examples from other age groups and women voices in the dataset. The dataset was collected by over 1,700 unique people,  16$\pm{14}$ utterances from each person with a median 14, and for each class, we have 407$\pm{31}$ unique speakers. This uniqueness and variety make this dataset suitable to be used in the speaker verification task too.

Finally, the dataset is divided into three parts: train, test, and continuous speech samples (CSS). In the test set, we select 300 samples in each class, considering that it contains at least 100 samples from speakers who are only available in the test part. Also, to evaluate the final models in continuous speeches, we prepared the CSS part. The CSS part consists of 1002 continuous 2-second speech samples, wherein each sample, the speaker utters one of the desired keywords in a continuous speech.

\subsection{Collecting and cleaning systems}
In order to gather desired words, we developed two telegram bots to collect OGG Vorbis audio files as crowdsourcing. The bots ask users to record and send the required six words based on the amount of data collected. Then users choose one of these words. Subsequently, the user was asked to utter the chosen word seven times, and each time with a different expression mood to obtain a diverse mood for each word. At any time, the user could go to the next word and is not forced to fill in all seven moods; however, results show that most of the users went to the next word after completing seven moods and then complete the next words in the same way. Another bot was also developed to determine the similarity of the audio files sent by users with famous Persian football reporters. To find out the most similar reporter to their voice, users must submit all 42 samples to this bot. This bot makes some fun and motivates people to send their voices because of its gamification idea.

There exists no time limit during data collection, and only users are asked to utter the keyword. The collected data was given to a cleaning system after they were converted to the \textit{wav} format. We train the cleaning system with 20000 samples cleaned by a human operator. To provide the required clean data, first, samples are trimmed systematically to 1.25-second based on more energy segment. Then, a human operator verifies and labels them. The cleaning system is based on the Res-8 architecture proposed in \cite{honk}. Then in the cleaning phase, for each audio file, by moving the 1.25-second sliding window with a stride of 0.1 second, 1.25-second audio parts are cropped. Next, among these samples, the sample with the highest probability of keyword occurrence is stored as the output of that audio file if it has a higher probability than a threshold value which is set to 0.97. If the audio file is less than 1.25 seconds, zero paddings are made equally to the beginning and end of the file to make the audio file at least 1.25-second. Then, three human operators clean the dataset, and we consider the intersection of each operator output samples as the final dataset. Table~\ref{tbl:fkd_info} indicates the number of samples of each keyword.

In order to improve the generalization and performance of any system using this dataset, we prepared 11 different noises, including white, pink, restaurant, exhibition, car, crowd cheering, running, subway, talking, doing the dishes, and babble. These noises have adequate diversities to simulate real situations and can be added to the samples during the training and test phases.

 \subsection{Synthesizing continuous speech}
 FKD consists of samples that contain only a single word like other datasets in this field. However, this type of datasets are not suitable for continuous speeches. Consequently, we need to convert samples into continuous ones. With a continuous speech synthesis method (CSSM), noncontinuous speech datasets can be converted to a proper dataset for continuous speeches.

To prepare synthesis continuous sample from $i^{th}$ sample $(\textbf{s}_i)$ with the sample rate of 16 KHz, first, we randomly select one of the continuous speech audio files $(\textbf{s}_g)$ from some football report files or any continuous speech file like movies, radio programs, etc.
\begin{equation*}
  \textbf{s}_g \in {\{background \ files\}}
\end{equation*}
Then, we select a two-second slice from the selected file as the background speech $\textbf{b}_g$.
\begin{equation*}
\textbf{b}_g = \textbf{s}_g[n, n+32000]
\end{equation*}
where $n$ is a random integer in the range of $[0,|\textbf{s}_g|-32000]$ and $|\textbf{s}_g|$ means the length of $\textbf{b}_g$. Therefore, two background and keyword windows according to equation~\ref{eq:w_kw} and~\ref{eq:w_bg} are created, respectively:
\begin{equation}
\textbf{w}_{kw}\left(j\right)=\frac{I_0\left(1.5\sqrt{1-\frac{4j^2}{\left(M-1\right)^2}}\right)}{I_0\left(1.5\right)}
\label{eq:w_kw}
\end{equation}
\begin{equation}
\textbf{w}_{bg}\left(j\right)=\ 1.05-\frac{I_0\left(2.5\sqrt{1-\frac{4j^2}{\left(M-1\right)^2}}\right)}{I_0\left(2.5\right)}
\label{eq:w_bg}
\end{equation}
where $I_0$ is the modified zeroth-order Bessel function. To makes the final output more natural, $\textbf{w}_{bg}$ has $0.125$ second zero-padding on both sides.
We randomly choose $k$ in the range $[bound, |\textbf{b}_g|-(bound+20000)]$ and multiply the background window by $\textbf{b}_g$ to prepare the desired background speech.
\begin{equation}
\textbf{b}^\prime_g[i] = \begin{cases}
\textbf{b}_g[i] & if\ i \notin [k,k+20000]\\
\textbf{b}_g[i]\times \textbf{w}_{bg}[i-k] &  o.w. \end{cases}
\end{equation}
Then, by multiplying the keyword window by sample and placing the result in the desired background speech, the final synthesized continuous speech is obtained $(\textbf{s}_s)$.
\begin{equation}
\textbf{s}_s[j] = \begin{cases}
\textbf{b}^\prime_g[i] & if\ i \notin [k+2000,k+18000]\\
\textbf{b}^\prime_g[i]+ \textbf{s}^\prime[i-(k+2000)] & o.w.\end{cases}
\end{equation}
where
\begin{equation*}
\textbf{s}^\prime[p] = \textbf{s}[p] \times \textbf{w}_{kw}[p]
\end{equation*}
 The steps of this process, along with the shape of the windows, are shown in Figure~\ref{fig:cssm}. As can be seen, the synthesized sample looks like a real continuous speech based on the graph. We also investigate the impact of using this method in improving the performance of our system in the continuous speech.

\begin{table}[h]\centering
\caption{The details of EfficientNet-A0 stages. Each row describes the number of stage alone with information about operator, kernel size, number of channels and number of layers.}

\label{tbl:effa0}
\begin{tabular}{c|cccc}
\hline
Stage  & Operator & Kernel size & \#Channels & \#Layers \\
\hline
1 & Conv & 3x3 & 16 & 1\\
2 & MBConv1 & 3x3 & 8 & 1\\
3 & MBConv6 & 5x5 & 16 & 2\\
4 & MBConv6 & 3x3 & 24 & 1\\
5 & MBConv6 & 3x3 & 32 & 2\\
6 & MBConv6 & 5x5 & 56 & 2\\
7 & MBConv6 & 3x3 & 96 & 2\\
8 & Conv, Pooling, FC & 1x1 & 384 & 1\\
\hline
\end{tabular}
\end{table}

\section{Evaluating the models}
Looking for a high-performance keyword spotting system with low latency and runtime, we used EfficientNet architecture \cite{eff}. The original article of this architecture mentioned that there exist eight models named B-0 to B-7 according to the compound scaling approach, in which the number of parameters has been added to the depth, width, and resolution of the input image, respectively. The result of these networks are significantly better than other existing structures such as ResNet, Inception, and MobileNet in cases with even fewer parameters. This network is powerful and effective in the field of speech processing too. As a consequence, we use this approach and scaling down the B0 model to achieve the proposed goal.

\begin{figure}[t]
  \centering
  \includegraphics[width=10cm, height= 12cm]{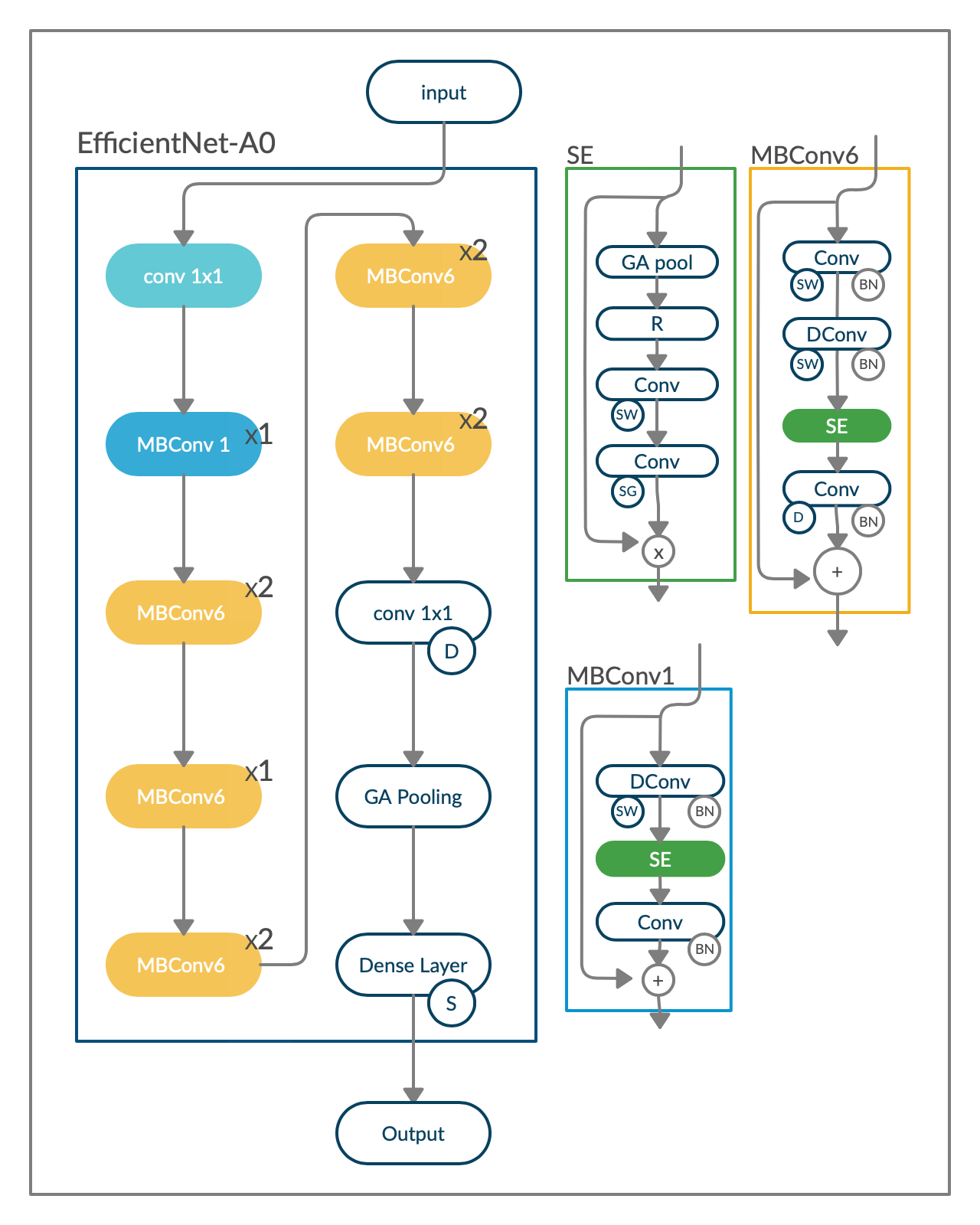}
  \caption{EfficientNet-A0 architecture. D, S, SW, SG, GA, R, and BN are abbreviations of Dropout, Softmax, Swiss, Sigmoid, Global Average, Reshape, and Batch Normalization, respectively.}
  \label{fig:effa0}
\end{figure}

It is desired to have a model with about 200000 to 250000 parameters to satisfy our problem's objectives. To do this, we could only change the depth and width of the network because the input size was appropriate due to its fixed and unchangeable time and features. Therefore, the goal is to reduce the model size as bellow, contrary to the method introduced in the EfficientNet's article:
\begin{equation}
\begin{split}
depth:\quad d = \alpha^\phi\\
width:\quad w = \beta^\phi\\
resolution:\quad r = \gamma^\phi \\
s.t.\quad\alpha.\beta^2.\gamma^2 \approx 0.05\pm 0.003
\end{split}
\end{equation}
where 0.05 is come from how much we want to reduce model's parameters which is $\frac{200000}{4000000} = 0.05$, and $0.003$ is the maximum bound that the final model can pass throw the limitation. This values is set based on 5\% of the number of Res15's parameters. we set $\gamma=1$, and select values in the range of $[0.25, 0.6]$ for $\alpha$ and $\beta$ parameters. Finally, it is required to train 75 models according to the existing conditions and grid search with the step size of $0.01$. To train and evaluate these models, only 100 random samples from each category are selected.
The architecture of EfficientNet-A0 is depicted in Figure~\ref{fig:effa0} and details about the stages, kernel size, and the number of channels are shown in Table~\ref{tbl:effa0}. We reduce the EfficientNet-B0 parameters from 4032595 to 238250 and obtain EfficientNet-A0. This reduction is achieved by decreasing the number of layers and channels.

In order to model input, Honk \cite{cnnsmall} procedure is executed. To this end, a band-pass filter of 20Hz/4kHz is applied to the input audio for noise reduction; then, 40 Mel-frequency cepstrum coefficients (MFCC) are extracted from each 30-milliseconds frame. Finally, to create two-dimensional input, this frame window is moved with steps of 10-milliseconds and the output frames are stacked. In order to have a generalized model, two approaches in the training phase are exploited. The first one is using a fast and novel SpecAugment method \cite{specaug} for data augmentation during the training phase. The frequency and time masks are applied on each batch sample with equal probabilities. In the second approach, during the training process, the proposed continuous speech synthesis method is employed to make each 1.25-second sample into a 2-second continuous speech. With this procedure, the trained model could see several continuous speeches with different backgrounds for each sample.

To make the proposed method usable in practical applications, we have considered two silence and unknown labels besides classes in the FKD. For silent samples, just a random 2-second file from one of the noises for each required sample is separated. Besides, two types of samples for the unknown class is considered during the training. Samples that are synthesized using the GSC test part, along with the samples which are just 2-second of the background files contaminated with additional noise.  Models are trained by minimizing a weighted cross-entropy loss function to mitigate the imbalance in the training dataset. Finally, to evaluate the proposed model, several available models with proper results in GSC have been tested. These models are Res8, Res15, and three convolutional methods trad-fpool13, tpool12, and one-srtide1 proposed in \cite{cnnsmall}.

\begin{table*}[!]\centering
\caption{Test and CSS accuracy (Acc.) of each model with 95\% confidence intervals (across five trials) on the FKD. The number of model parameters is normalized based on the size of EfficientNet-A0 model, which is 238250 parameters.}
\label{tbl:result}
\begin{tabular}{lccccl}
\hline
\multicolumn{1}{c}{\textbf{Model}} & \textbf{\# Parameters} &\textbf{\# GFLOPS} & \textbf{Test Acc. (\%)} & \textbf{CSS Acc. (\%)} &  \\
\hline
Res8                               & 0.44x	&0.08                  & 94.38$\pm{0.451}$                 & 75.54$\pm{0.358}$                   &  \\
Res15                              & 0.95x 	&1.93                 & 94.85$\pm{0.349}$              & 74.85$\pm{0.312}$                   &  \\
Res26                              & 1.75x 	&0.89                 & \textbf{95.88}$\pm{\textbf{0.253}}$           & 69.46$\pm{0.263}$                  &  \\
trad-fpool3                        & 0.51x 	&0.15                & 74.46$\pm{0.403}$                    & 44.01$\pm{0.308}$                   &  \\
tpool2                             & 9.15x	&0.24                  & 89.91$\pm{0.468}$                    & 61.17$\pm{0.355}$                   &  \\
one-srtide1                        & 4.4x	&0.58                  & 78.60$\pm{0.832}$                    & 51.99$\pm{0.538}$                   &  \\
EfficientNet-A0                    & 1x	&0.01                     & 94.56$\pm{0.302}$                    & 73.25$\pm{0.212}$                   &  \\
EfficientNet-A0 + SA               & 1x	&0.01                     & 94.83$\pm{0.377}$                    & 76.22$\pm{0.231}$                   &  \\
EfficientNet-A0 + SA + TL          & 1x	&0.01                     & 95.83$\pm{0.401}$                    & \textbf{76.84}$\pm{\textbf{0.240}}$         &\\
EfficientNet-A0  (without CSSM)                 & 1x	&0.01                     & -                    & 63.45$\pm{0.412}$                   &  \\
\hline
\end{tabular}
\end{table*}

\section{Experiments}
This section provides an overview of the general configuration, results, and a description of the experiments performed.

\subsection{General configuration}
Honk open-source project is used as our primary method and is modified it based on our requirements. The implementation is available on the git-hub\footnote{https://github.com/AmirmohammadRostami/KeywordsSpotting-EfficientNet-A0}. To do the experiments, as mentioned in the previous section, we employed the eight models whose performances are significant on the GSC, along with new EfficientNet architecture, which to the best of our knowledge, are used for the first time in this field. To evaluate the models in a fair conditions, constant values are set for common parameters. In other words, we set epoch number 40; batch-size 128; Adam optimizer with $\beta$=(0.9,0.999); schedule learning rate based on cosine annealing with $\eta_{max}$=0.1  and $\eta_{min}$=0.01 ; and the probability value of 0.5 for augmentation functions including frequency and time masking with F=5 and T=8; and finally, we set noise multiplier N=0.12 based on the GSC.

To avoid over-fitting, 20\% of training data is considered to evaluate the models at each epoch. We choose the model with the best result on evaluation data in the training phase as a final model. Then for the final evaluation, test data is used.

\subsection{Results}

Table~\ref{tbl:result} shows the performance of different models on the FKD. As shown in this table, the EfficientNet-A0 model with spec augmentation (SA) and transfer learning (TL) has the best performance in continuous speech samples (CSS). Besides, in the test part, all Resnet models and EfficientNet-A0 have nearly the same performances. We also try Resnet-Narrow models, but their results are inferior, showing us increase the number of channels in each layer is more useful than increasing the model depth.

In order to evaluate the complexity, two metrics, the number of parameters and floating-point operations (FLOPS), which indicate the model's complexity and execution time during the inference phase have been used\cite{flops}\cite{param}. As it can be observed in Table~\ref{tbl:result}, the number of parameters of EfficientNet-A0 is much less than other traditional convolutional methods and is approximately equal to the number of parameters of ResNets. However, GFLOPS (G is for Giga) required in the inference phase of EfficientNet-A0 is significantly fewer, which indicates that this model outperforms all models with FLOPS equal to 7439100.

Also, we investigate EfficientNet-A0 with two approaches SA and TL. In transfer learning models, the model is first trained on the GSC dataset and then its pre-trained weights are used as the initial weight. Results show that this mechanism works well with the GSC and FKD and achive the best result in the CSS part. By taking the advatage of spec augmentation method, we improve the model generalization in the test phase especially in the CSS part. Finally, we evaluated the CSSM impact by training EfficientNet-A0 without using this approach in the training phase. The last row of Table~\ref{tbl:result} shows that this model has 63.45$\pm{0.412}\%$ accuracy on the CSS part, indicating that using the CSSM models performs better in continuous speeches.

\section{Conclusion}
In this paper, we presented a new keyword spotting dataset for football event detection in \textit{Persian}. This dataset was collected with crowdsourcing and is available as an open-source project. It contains 18 classes, which are football events keywords. The FKD contains about 31000 samples and is divided into three parts including the train, test, and continuous speech samples. Moreover, the continuous speech synthesis method is employed to convert samples containing only a single keyword to continuous speech samples in order to achive models working better with continuous speeches.

Finally, we used the compound scaling approach on EfficientNet-B0 architecture to create our proposed architecture EfficientNet-A0. This architecture with specAugmentation and transfer learning achieved the best result for the CSS part by 76.84$\pm{0.240}\%$ accuracy and also 95.83$\pm{0.401}\%$ accuracy in the test part. The result shows that ResNets and EfficientNet-A0 outperform traditional convolutional methods. Also, the EfficientNet-A0 has only 0.01 GFPLOS, which is a long way from other models, especially Res26, which is the closest in terms of accuracy. This point, along with the number of parameters, indicates the efficiency of this model in real-time applications.

\bibliographystyle{unsrt}  
\bibliography{references}  

\begin{thebibliography}{10}

\bibitem{eff}
Mingxing Tan and Quoc~V. Le.
\newblock Efficientnet: Rethinking model scaling for convolutional neural
  networks.
\newblock {\em CoRR}, abs/1905.11946, 2019.

\bibitem{effspeech}
Qidong Lu, Yingying Li, Zhiliang Qin, Xiaowei Liu, and Yun Xie.
\newblock Speech recognition using efficientnet.
\newblock New York, NY, USA, 2020. Association for Computing Machinery.

\bibitem{cnn2}
H.~Salehinejad, J.~Barfett, P.~Aarabi, S.~Valaee, Errol Colak, B.~Gray, and
  T.~Dowdell.
\newblock A convolutional neural network for search term detection-a draft.
\newblock {\em ArXiv}, abs/1708.02238, 2017.

\bibitem{cnn1}
H.~{Lim}, Y.~{Kim}, Y.~{Kim}, and H.~{Kim}.
\newblock Cnn-based bottleneck feature for noise robust query-by-example spoken
  term detection.
\newblock In {\em 2017 Asia-Pacific Signal and Information Processing
  Association Annual Summit and Conference (APSIPA ASC)}, pages 1278--1281,
  2017.

\bibitem{bbn}
T.~{Alumäe}, D.~{Karakos}, W.~{Hartmann}, R.~{Hsiao}, L.~{Zhang}, L.~{Nguyen},
  S.~{Tsakalidis}, and R.~{Schwartz}.
\newblock The 2016 bbn georgian telephone speech keyword spotting system.
\newblock In {\em 2017 IEEE International Conference on Acoustics, Speech and
  Signal Processing (ICASSP)}, pages 5755--5759, 2017.

\bibitem{honk}
R.~{Tang} and J.~{Lin}.
\newblock Deep residual learning for small-footprint keyword spotting.
\newblock In {\em 2018 IEEE International Conference on Acoustics, Speech and
  Signal Processing (ICASSP)}, pages 5484--5488, 2018.

\bibitem{cnnsmall}
Tara Sainath and Carolina Parada.
\newblock Convolutional neural networks for small-footprint keyword spotting.
\newblock In {\em Interspeech}, 2015.

\bibitem{DNN1}
G.~{Chen}, C.~{Parada}, and G.~{Heigold}.
\newblock Small-footprint keyword spotting using deep neural networks.
\newblock In {\em 2014 IEEE International Conference on Acoustics, Speech and
  Signal Processing (ICASSP)}, pages 4087--4091, 2014.

\bibitem{DNN2}
R.~{Prabhavalkar}, R.~{Alvarez}, C.~{Parada}, P.~{Nakkiran}, and T.~N.
  {Sainath}.
\newblock Automatic gain control and multi-style training for robust
  small-footprint keyword spotting with deep neural networks.
\newblock In {\em 2015 IEEE International Conference on Acoustics, Speech and
  Signal Processing (ICASSP)}, pages 4704--4708, 2015.

\bibitem{DNN3}
George Tucker, Minhua Wu, Ming Sun, Sankaran Panchapagesan, Gengshen Fu, and
  Shiv Vitaladevuni.
\newblock Model compression applied to small-footprint keyword spotting.
\newblock pages 1878--1882, 09 2016.

\bibitem{rnn2}
Ming Sun, Anirudh Raju, George Tucker, Sankaran Panchapagesan, Gengshen Fu,
  Arindam Mandal, Spyros Matsoukas, Nikko Strom, and Shiv Vitaladevuni.
\newblock Max-pooling loss training of long short-term memory networks for
  small-footprint keyword spotting.
\newblock pages 474--480, 12 2016.

\bibitem{tfc}
Taejun Kim and Juhan Nam.
\newblock Temporal feedback convolutional recurrent neural networks for keyword
  spotting, 10 2019.

\bibitem{crnn}
Sercan Arik, Markus Kliegl, Rewon Child, Joel Hestness, Andrew Gibiansky, Chris
  Fougner, Ryan Prenger, and Adam Coates.
\newblock Convolutional recurrent neural networks for small-footprint keyword
  spotting.
\newblock 03 2017.

\bibitem{crnn2}
M.~{Zeng} and N.~{Xiao}.
\newblock Effective combination of densenet and bilstm for keyword spotting.
\newblock {\em IEEE Access}, 7:10767--10775, 2019.

\bibitem{hmm}
R.~C. {Rose} and D.~B. {Paul}.
\newblock A hidden markov model based keyword recognition system.
\newblock In {\em International Conference on Acoustics, Speech, and Signal
  Processing}, pages 129--132 vol.1, 1990.

\bibitem{hyb}
{Jiazhi Ou}, {Kaijiang Chen}, {Xiuping Wang}, and {Zongge Li}.
\newblock Utterance verification of short keywords using hybrid
  neural-network/hmm approach.
\newblock In {\em 2001 International Conferences on Info-Tech and Info-Net.
  Proceedings (Cat. No.01EX479)}, volume~2, pages 671--676 vol.2, 2001.

\bibitem{msr}
A.~{Garcia} and H.~{Gish}.
\newblock Keyword spotting of arbitrary words using minimal speech resources.
\newblock In {\em 2006 IEEE International Conference on Acoustics Speech and
  Signal Processing Proceedings}, volume~1, pages I--I, 2006.

\bibitem{med}
Kartik Audhkhasi and Ashish Verma.
\newblock Keyword search using modified minimum edit distance measure.
\newblock volume~4, pages IV--929, 05 2007.

\bibitem{scd}
Pete Warden.
\newblock Speech commands: A dataset for limited-vocabulary speech recognition.
\newblock {\em ArXiv}, abs/1804.03209, 2018.

\bibitem{mozila}
Rosana Ardila, Megan Branson, Kelly Davis, Michael Henretty, M.~Kohler, Josh
  Meyer, Reuben Morais, Lindsay Saunders, Francis~M. Tyers, and Gregor Weber.
\newblock Common voice: A massively-multilingual speech corpus.
\newblock In {\em LREC}, 2020.

\bibitem{deepmine}
H.~{Zeinali}, L.~{Burget}, and J.~H. {Černocký}.
\newblock A multi purpose and large scale speech corpus in persian and english
  for speaker and speech recognition: The deepmine database.
\newblock In {\em 2019 IEEE Automatic Speech Recognition and Understanding
  Workshop (ASRU)}, pages 397--402, 2019.

\bibitem{specaug}
Daniel~S. Park, William Chan, Yu~Zhang, Chung-Cheng Chiu, Barret Zoph,
  Ekin~Dogus Cubuk, and Quoc~V. Le.
\newblock Specaugment: A simple augmentation method for automatic speech
  recognition.
\newblock In {\em INTERSPEECH}, 2019.

\bibitem{flops}
Seungwoo Choi, Seokjun Seo, Beomjun Shin, Hyeongmin Byun, Martin Kersner,
  Beomsu Kim, Dongyoung Kim, and Sungjoo Ha.
\newblock Temporal convolution for real-time keyword spotting on mobile
  devices.
\newblock 04 2019.

\bibitem{param}
R.~{Tang}, W.~{Wang}, Z.~{Tu}, and J.~{Lin}.
\newblock An experimental analysis of the power consumption of convolutional
  neural networks for keyword spotting.
\newblock In {\em 2018 IEEE International Conference on Acoustics, Speech and
  Signal Processing (ICASSP)}, pages 5479--5483, 2018.

\end{thebibliography}

\end{document}